\documentclass[11pt]{article}
\usepackage{moriond,epsfig}

\def\be{\begin{equation}}
\def\ee{\end{equation}}
\def\bea{\begin{eqnarray}}
\def\eea{\end{eqnarray}}

\newcommand{\bsmumu}{$B_s\to\mu^+\mu^-$}

\begin{document}
\vspace{-3cm}
\title{Overview of flavour physics with focus on the MSSM and 2HDMs}

\author{ANDREAS CRIVELLIN}

\address{CERN Theory Division, CH-1211 Geneva 23, Switzerland}

\maketitle\abstracts{In these proceedings we give a concise review of some selected flavour-violation processes and their implications for two-Higgs-doublet models (2HDMs) and the MSSM. The processes under investigation are $\Delta F=2$ processes, \bsmumu, $b\to s\gamma$, and tauonic $B$ decays. For each process we show the impact on the models.}

\section{Introduction}

In recent years flavour physics has been one of the most active and
fastest developing fields in high energy physics. Numerous new
experiments were carried out but almost all of them reported result in agreement with the Standard Model (SM) predictions. There are only a few exceptions like the anomaly in the anomalous magnetic moment of the muon or recently the deviations from the SM predictions in tauonic $B$ decays \cite{BaBar:2012xj} and $B\to K^*\mu^+\mu^-$ \cite{Aaij:2013qta}.

The extensive set of measurements available for rare decays puts
strong constraints on the flavour structure of physics beyond the
Standard Model, in particular, on the flavour- and CP- violating parameters of the Minimal Supersymmetric
Standard Model (MSSM) (see for example \cite{Altmannshofer:2009ne} for an overview) or the two-Higgs doublet model (2HDM) \cite{Crivellin:2013wna}. 

The SM has only one Higgs doublet and in a 2HDM (which is the decoupling limit of the MSSM) we introduce a second Higgs doublet and obtain four additional physical Higgs particles (in the case of a CP conserving Higgs potential): the neutral CP-even Higgs $H^0$, a neutral CP-odd Higgs $A^0$ and the two charged Higgses $H^{\pm}$. In addition, if we allow for a generic flavour structure we have the non-holomorphic couplings which couple up (down) quarks to the down (up) type Higgs doublet: $\bar u_f \epsilon^u_{fi}u_i H^d$ and $d_f \epsilon^d_{fi}d_i H^u$ where $\epsilon^q_{ij}$ parametrizes the completely flavour-chaining neutral currents.

In the MSSM at tree-level $\epsilon^q_{ij}=0$ (which corresponds to the 2HDM of type II) and flavour changing neutral Higgs couplings are absent. However, these couplings are generated at the loop level. The resulting expressions are non-decoupling and depend only on the ratios of SUSY parameters (for a complete one-loop analysis see \cite{Crivellin:2011jt} and for the 2-loop SQCD corrections \cite{Crivellin:2012zz}).

\section{Selected flavour-processes and their implications}

\subsection{$\Delta F=2$ processes}

$\Delta F=2$ processes are still one of the most constraining processes for NP (see \cite{Lenz:2010gu} for an overview on $B_q-\bar B_q$ mixing) since they scale like $\delta^2/\Lambda^2$ while the other flavour observables scale like $\delta/\Lambda^2$. Here $\delta$ stands for a generic flavour violating parameter and $\Lambda$ is the scale of NP. Especially the constraints from Kaon and D mixing are very stringent. They can be used for example to constrain the mass splitting of left-handed squarks in the MSSM \cite{Crivellin:2010ys} (see left plot of Fig.~\ref{fig:Bstomumu}).

\subsection{$B_q\to\mu^+\mu^-$}

Thanks to LHCb and CMS \cite{Aaij:2012nna} we know the branching ration for $B_s\to\mu^+\mu^-$ now rather precisely and also the SM prediction has been improved recently \cite{Bobeth:2013uxa}:
\begin{equation}
	{\rm Br}{\left[ {{B_s} \to {\mu ^ + }{\mu ^ - }} \right]_{\exp }} = \left( {2.9 \pm 0.7} \right) \times {10^{ - 9}}\,,\qquad {\rm Br}{\left[ {{B_s} \to \mu \mu } \right]_{SM}} = \left( {3.65 \pm 0.23} \right) \times {10^{ - 9}}\,.
\end{equation}
Due to the good agreement with the SM we can place stringent bounds on models of NP, especially if they have sizable flavour-changing scalar currents like the generic 2HDM or the MSSM at large $\tan\beta$. In the middle plot of Fig.~\ref{fig:Bstomumu} we show the constraints on the 2HDM parameter $\epsilon^d_{23,32}$ which generate $B_s\to\mu^+\mu^-$ via a tree-level Higgs exchange.

While the experimental bounds on $B_d\to\mu^+\mu^-$ are still weaker due to the further CKM suppressed SM contribution LHCb will further improve experimental limit in the future. Also here stringent limits on $\epsilon^d_{13,31}$ can be obtained and similarly $K_L\to\mu^+\mu^-$ and $D\to\mu^+\mu^-$ but bounds on $\epsilon^q_{12,21}$. In summary, neural meson decays to muons constrain all flavour-chaning elements $\epsilon^d_{ij}$ and $\epsilon^u_{12,21}$ stringently.

\subsection{$b\to q\gamma$}

\begin{figure}[t]
\centering
\includegraphics[width=0.3\textwidth]{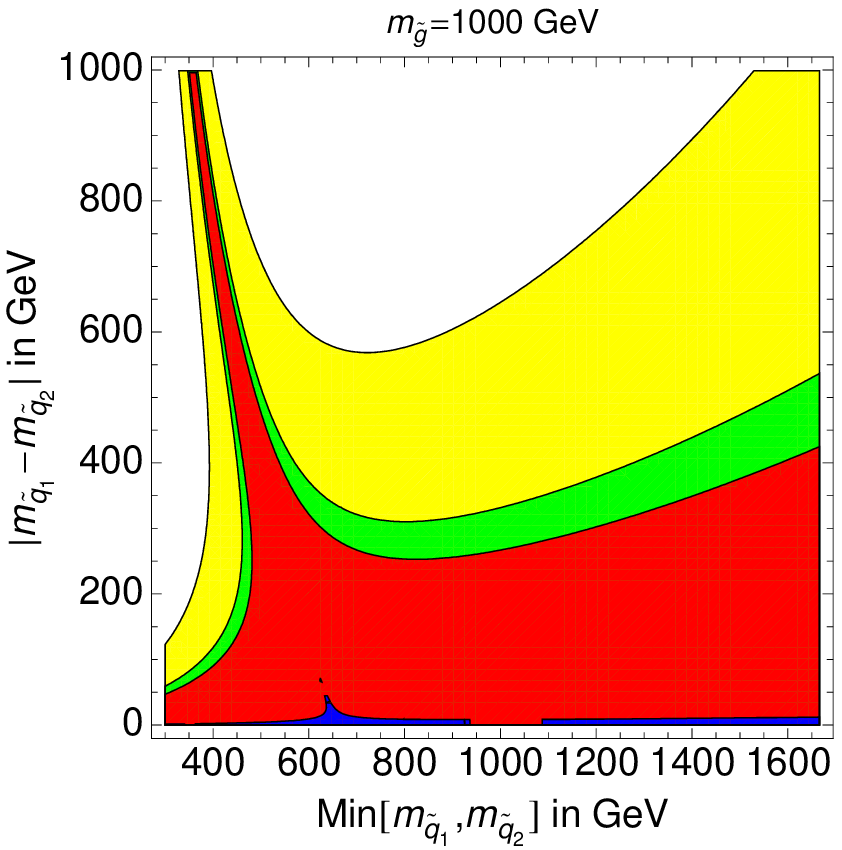}~~~~~
\includegraphics[width=0.3\textwidth]{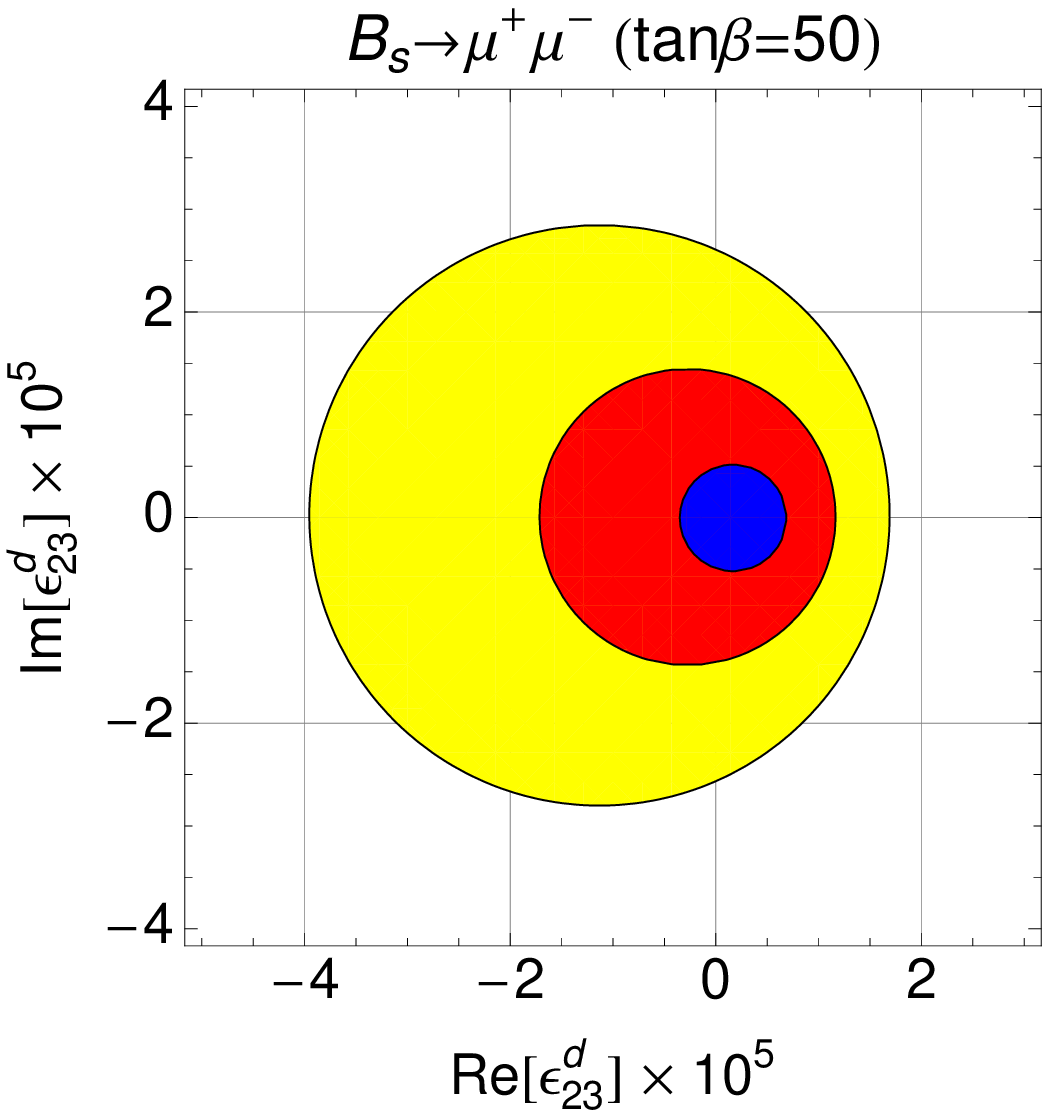}
\includegraphics[width=0.334\textwidth]{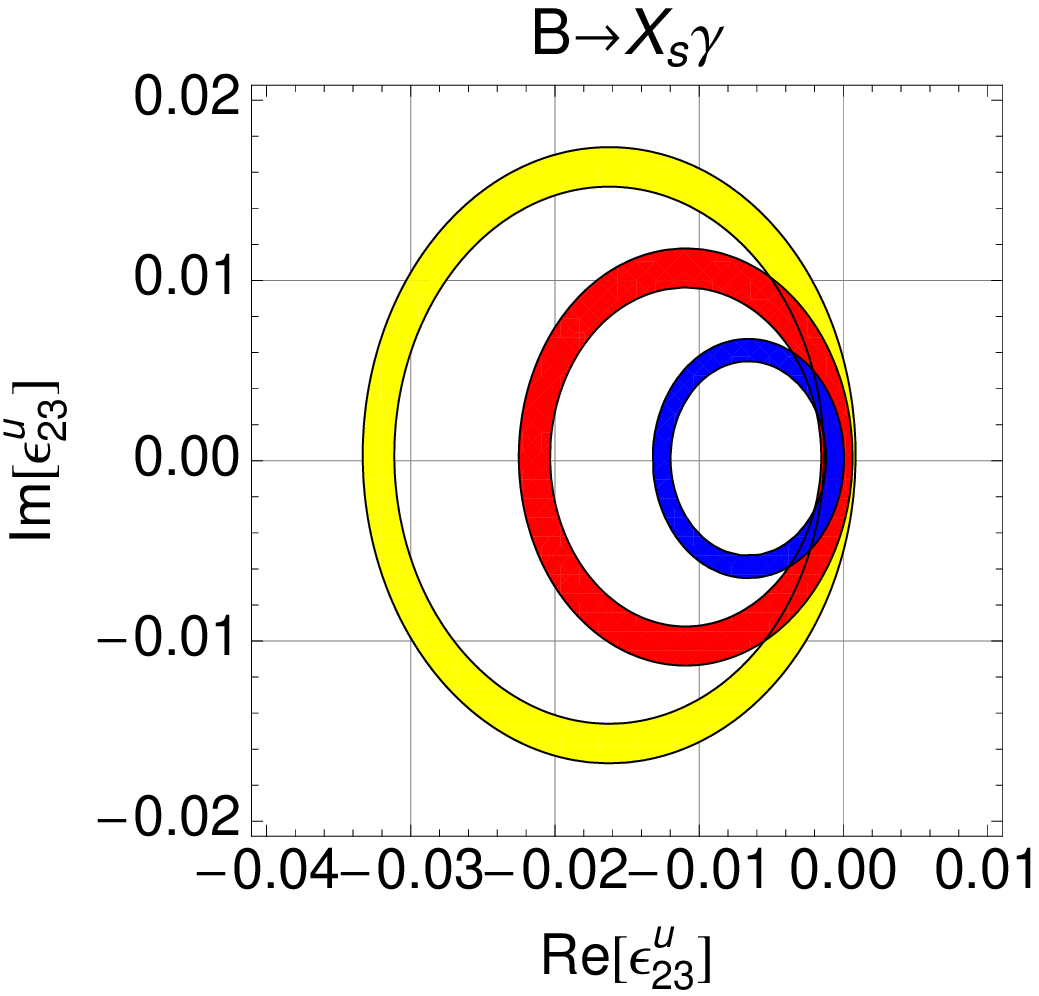}
\caption{Left: Allowed mass splitting between the first two generations of left-handed squarks for different gluino masses for $M_2=(\alpha_2/ \alpha_s) m_{\tilde g}\cong 0.35$. Yellow (lightest) corresponds to the maximally allowed mass splitting assuming an intermediate alignment of $m^2_{\tilde{q}}$ with $Y_u^{\dagger}Y_u$ and $Y_d^{\dagger}Y_d$. The green (red) region is the allowed range assuming an diagonal up (down) squark mass matrix. The blue (darkest) area is the minimal region allowed in which the off-diagonal element carries a maximal phase. Middle: Allowed regions in the complex $\epsilon^{d}_{23}$--plane from $B_s\to\mu^+\mu^-$ for $\tan\beta=50$ and $m_{H}=700\mathrm{~GeV}$ (yellow), $m_{H}=500\mathrm{~GeV}$ (red) and $m_{H}=300\mathrm{~GeV}$ (blue). Note that the allowed regions for $\epsilon^{d}_{32}$--plane are not full circles because in this case a suppression of ${\cal B}\left[B_{s}\to\mu^+\mu^-\right]$ below the experimental lower bound is possible. Right: Allowed regions for $\epsilon^{u}_{23}$ from $ B \to X_{s} \gamma$, obtained by adding the $2\,\sigma$ experimental error and theoretical uncertainty linear for $\tan\beta=50$ and $m_{H}=700 \, \mathrm{ GeV}$ (yellow), $m_{H}=500\, \mathrm{ GeV}$ (red) and  $m_{H}=300 \,\mathrm{ GeV}$ (blue). }
\label{fig:Bstomumu}
\end{figure}

Concerning the radiative $B$ decays $b\to s\gamma$ and $b\to d\gamma$ the current experimental values and theoretical predictions are given by:
\begin{equation}
\begin{array}{l}
 {\rm Br}{\left[ {b \to s\gamma } \right]_{\exp }} = \left( {3.43 \pm 0.21 \pm 0.07} \right) \times {10^{ - 4}}\,,\qquad
 {\rm Br}{\left[ {b \to s\gamma } \right]_{SM}} = \left( {3.13 \pm 0.22} \right) \times {10^{ - 4}}\,, \\ 
 {\rm Br}{\left[ {b \to d\gamma } \right]_{\exp }} = \left( {1.41 \pm 0.57} \right) \times {10^{ - 5}} \,,\qquad\qquad\;\;\;\,
 {\rm Br}{\left[ {b \to d\gamma } \right]_{SM}} = 1.54_{ - 0.31}^{ + 0.26} \times {10^{ - 5}} \,.\\ 
 \end{array}
\end{equation}
Again, we observe a good agreement between theory predictions\footnote{The SM prediction for $b \to d\gamma$ is taken from \cite{Crivellin:2011ba} while the value for $b\to s\gamma$ is a preliminary result  presented in Portoroz 2013 by Mikolaj Misiak.} and experiment. $b\to s\gamma$ can for example be used to put bounds on $\epsilon^u_{23}$ originating from charged Higgs loop contributions. The results are shown in the right plot of Fig.~\ref{fig:Bstomumu}. Similar constrains apply for $\epsilon^u_{13}$ from $b\to d\gamma$.

\subsection{Tauonic $B$ decays}

Tauonic $B$-meson decays are an excellent probe of new physics: they test lepton flavor universality satisfied in the SM and are sensitive to new particles which couple proportionally to the mass of the involved particles (e.g. Higgs bosons) due to the heavy $\tau$ lepton involved. Recently, the BABAR collaboration performed an analysis of the semileptonic $B$ decays $B\to D\tau\nu$ and $B\to D^*\tau\nu$ using the full available data set \cite{BaBar:2012xj}. They find for the ratios
\begin{equation}
{\cal R}(D^{(*)})\,=\,{\cal B}(B\to D^{(*)} \tau \nu)/{\cal B}(B\to D^{(*)} \ell \nu)\,,
\end{equation}
the following results:
\begin{eqnarray}
{\cal R}(D)\,=\,0.440\pm0.058\pm0.042  \,,\qquad{\cal R}(D^*)\,=\,0.332\pm0.024\pm0.018\,.
\end{eqnarray}
Here the first error is statistical and the second one is systematic. Comparing these measurements to the SM predictions
\begin{eqnarray}
{\cal R}_{\rm SM}(D)\,=\,0.297\pm0.017 \,, \qquad{\cal R}_{\rm SM}(D^*) \,=\,0.252\pm0.003 \,,
\end{eqnarray}
we see that there is a discrepancy of 2.2\,$\sigma$ for $\cal{R}(D)$ and 2.7\,$\sigma$ for $\cal{R}(D^*)$ and combining them gives a $3.4\, \sigma$ deviation from the SM~\cite{BaBar:2012xj}. This evidence for new physics in $B$-meson decays to taus is further supported by the measurement of ${\cal B}[B\to \tau\nu]=(1.15\pm0.23)\times 10^{-4}$ which disagrees with by $1.6\, \sigma$ higher than the SM prediction using $V_{ub}$ from a global fit of the CKM matrix \cite{Charles:2004jd}.

A natural possibility to explain these enhancements compared to the SM prediction is a charged scalar particle which couples proportionally to the masses of the fermions involved in the interaction: a charged Higgs boson. A charged Higgs affects $B\to \tau\nu$~\cite{Hou:1992sy}, $B\to D\tau\nu$ and $B\to D^*\tau\nu$~\cite{Tanaka:1994ay}. In a 2HDM of type II (with MSSM like Higgs potential) the only free additional parameters are $\tan\beta=v_u/v_d$ (the ratio of the two vacuum expectation values) and the charged Higgs mass $m_{H^\pm}$ (the heavy CP even Higgs mass $m_{H^0}$ and the CP odd Higgs mass $m_{A^0}$ can be expressed in terms of the charged Higgs mass and differ only by electroweak corrections). In this setup the charged Higgs contribution to $B\to \tau\nu$ interferes necessarily destructively with the SM contribution\cite{Hou:1992sy}. Thus, an enhancement of $\cal B\left[B\to \tau\nu\right]$ is only possible if the absolute value of the charged Higgs contribution is bigger than two times the SM one\footnote{Another possibility to explain $B\to \tau\nu$ is the introduction of a right-handed $W$-coupling \cite{Crivellin:2009sd}.}. Furthermore, a 2HDM of type II cannot explain $\cal{R}(D)$ and $\cal{R}(D^*)$ simultaneously \cite{BaBar:2012xj}.

\begin{figure}[t]
\centering
\includegraphics[width=0.3\textwidth]{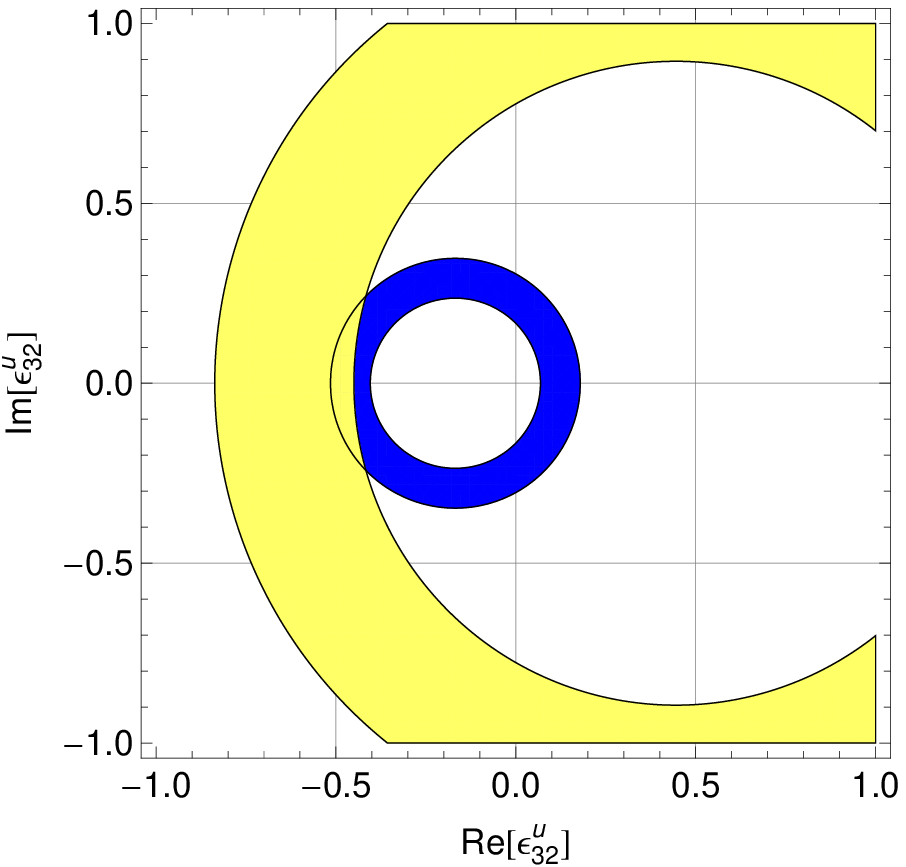}
\includegraphics[width=0.31\textwidth]{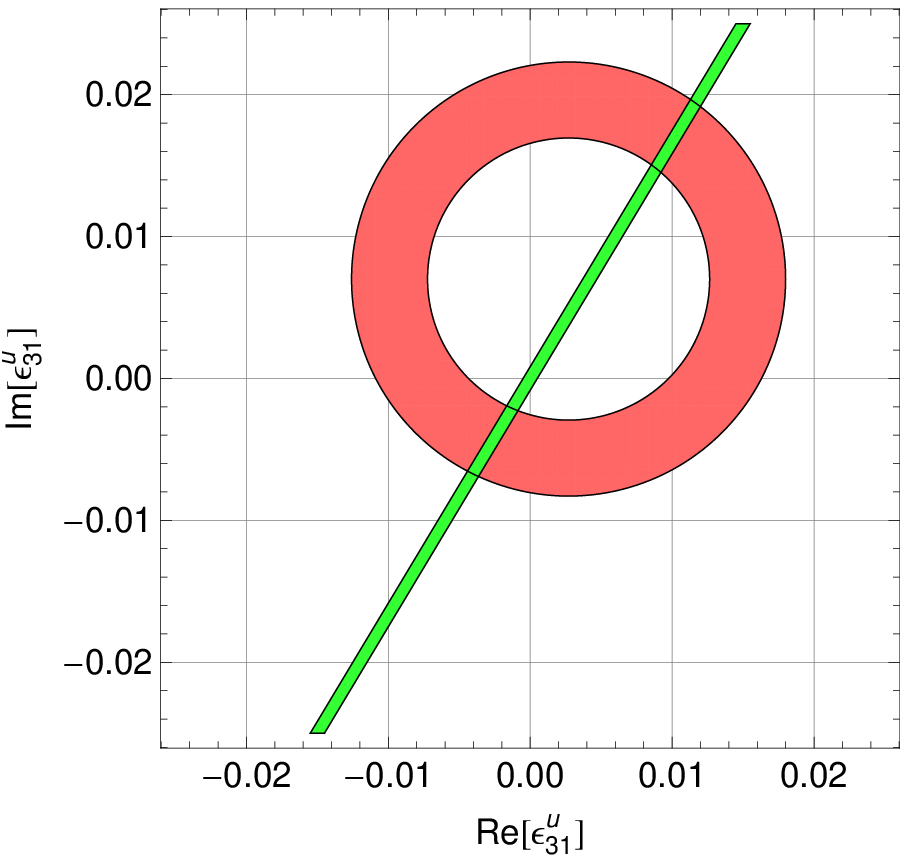}
\includegraphics[width=0.31\textwidth]{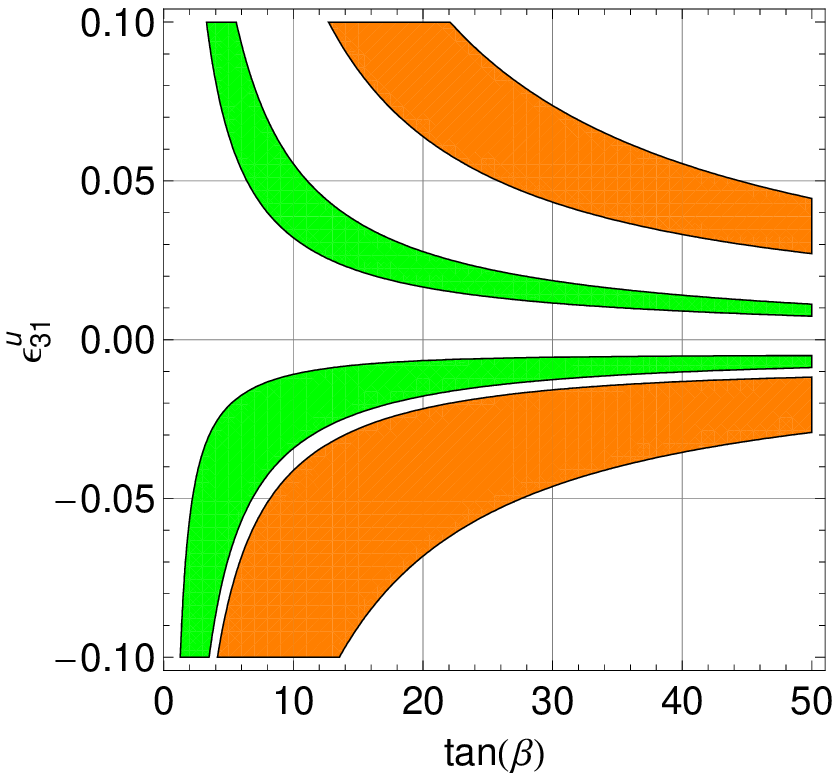}
\caption{Left: Allowed regions in the complex $\epsilon^u_{32}$-plane from $\cal{R}(D)$ (blue) and $\cal{R}(D^*)$ (yellow) for $\tan\beta=50$ and $m_H=500$~GeV. Middle:  Allowed regions in the complex $\epsilon^u_{31}$-plane from $B\to \tau\nu$. Right:  Allowed regions in the $\tan\beta$--$\epsilon^u_{31}$ plane from $B\to \tau\nu$ for real values of $\epsilon^u_{31}$ and $m_H=400$~GeV (green), $m_H=800$~GeV (orange). The scaling of the allowed region for $\epsilon^u_{32}$ with $\tan\beta$ and $m_H$ is the same as for $\epsilon^u_{31}$. $\epsilon^u_{32}$ and $\epsilon^u_{31}$ are given at the matching scale $m_H$. \label{2HDMIII}}
\end{figure}

As we found before, all $\epsilon^d_{ij}$ and $\epsilon^u_{13,23}$ are stringently constrained from FCNC processes in the down sector and only $\epsilon^u_{31}$ ($\epsilon^u_{32}$) significantly effects $B\to \tau\nu$ ($\cal{R}(D)$ and $\cal{R}(D^*)$) without any suppression by small CKM elements. Furthermore, since flavor-changing top-to-up (or charm) transitions are not measured with sufficient accuracy, we can only constrain these elements from charged Higgs-induced FCNCs in the down sector. However, since in this case an up (charm) quark always propagates inside the loop, the contribution is suppressed by the small Yukawa couplings of the up-down-Higgs (charm-strange-Higgs) vertex involved in the corresponding diagrams. Thus, the constraints from FCNC processes are weak, and $\epsilon^u_{32,31}$ can be sizable. 
Indeed, it turns out that by using $\epsilon^u_{32,31}$ we can explain $\cal{R}(D^*)$ and $\cal{R}(D)$ simultaneously \cite{Crivellin:2012ye}. In Fig.~\ref{2HDMIII} we see the allowed region in the complex $\epsilon^u_{32}$-plane, which gives the correct values for $\cal{R}(D)$ and $\cal{R}(D^*)$ within the $1\, \sigma$ uncertainties for $\tan\beta=50$ and $M_H=500$~GeV. Similarly, $B\to \tau\nu$ can be explained by using $\epsilon^u_{31}$.

\section*{Acknowledgments}

I thank the organizers, especially Maria Krawczyk, for the invitation and the possibility to present these results. This work is supported by a Marie Curie Intra-European Fellowship of the European Community's 7th Framework Programme under contract number (PIEF-GA-2012-326948).

\end{document}